\def\year{2020}\relax
\documentclass[letterpaper]{article} % DO NOT CHANGE THIS
\usepackage{aaai20}  % DO NOT CHANGE THIS
\usepackage{times}  % DO NOT CHANGE THIS
\usepackage{helvet} % DO NOT CHANGE THIS
\usepackage{courier}  % DO NOT CHANGE THIS
\usepackage[hyphens]{url}  % DO NOT CHANGE THIS
\usepackage{graphicx} % DO NOT CHANGE THIS
\usepackage{graphicx}  % DO NOT CHANGE THIS
\usepackage{algorithm}
\usepackage[noend]{algorithmic}
\usepackage{amsmath}
\usepackage{amsthm}
\usepackage{amsfonts}
\usepackage{amssymb}
\usepackage{comment}
\usepackage{enumitem}
\usepackage{graphicx}
\usepackage{listings}
\usepackage{makeidx}  % allows for indexgeneration
\usepackage{subcaption}
\usepackage{url}
\usepackage{multicol}
\usepackage{multirow}

\usepackage[graphicx]{realboxes}
\usepackage{tikz}
\tikzset{elliptic state/.style={draw,ellipse}}
\usetikzlibrary{shapes,shapes.geometric,arrows,fit,calc,positioning,automata,}
\usetikzlibrary{automata,positioning}
\usepackage{tikz-qtree}
\renewcommand*{\L}{\mathcal{L}}

\newcommand{\ltlf}{\mathsf{LTLf}}
\newcommand{\ltl}{\mathsf{LTL}}

\newcommand*{\splitformulalist}{\mathsf{splitFormulaList}}

\newcommand*{\dfasizeminheap}{\mathsf{dfaStateNumMinHeap}}
\newcommand*{\dfasymsizeminheap}{\mathsf{transNodeNumMinHeap}}

\newcommand*{\pop}{\mathsf{pop}}
\newcommand*{\push}{\mathsf{push}}
\newcommand*{\size}{\mathsf{size}}

\newcommand*{\explicitminimization}{\mathsf{explicitMinimal}}

\newcommand*{\makesymbolic}{\mathsf{makeSymbolic}}

\newcommand{\threshDFA}{t_1}
\newcommand{\threshProd}{t_2}

\newcommand*{\mona}{\mathsf{Mona}}
\newcommand*{\spot}{\mathsf{Spot}}
\newcommand*{\syft}{\mathsf{Syft}}

\newcommand*{\strix}{\mathsf{Strix}}
\newcommand*{\acaciaplus}{\mathsf{Acacia+}}

\newcommand{\lisa}{\mathsf{Lisa}}
\newcommand{\lisasynt}{\mathsf{LisaSynt}}

\newtheorem{theorem}{Theorem}

\newtheorem{definition}{Definition}

\newcommand{\ltlfU}{\mathsf{U}}
\newcommand{\ltlfX}{\mathsf{X}}
\newcommand{\ltlfNeg}{\neg}
\newcommand{\ltlfG}{\mathsf{G}}
\newcommand{\ltlfF}{\mathsf{F}}

\newcommand{\ltlloop}{\mathsf{loop}}

\newcommand{\alphabet}{\Sigma}
\newcommand{\states}{S}
\newcommand{\trans}{\Delta}
\newcommand{\init}{\iota}
\newcommand{\finals}{F}
\newcommand{\dfa}{D}
\newcommand{\ap}{\mathsf{Prop}}
\newcommand{\varset}[1]{\mathcal{#1}}

\newcommand{\partition}{\mathsf{Part}}

\newcommand{\sdfa}{\mathcal{D}}
\newcommand{\strans}{\mathcal{T}}
\newcommand{\sinit}{\mathcal{S}}
\newcommand{\sfinals}{\mathcal{F}}

\newcommand{\ZVar}{\varset{Z}}

\newcommand{\apI}{\varset{I}}
\newcommand{\apO}{\varset{O}}
\newcommand{\G}{\varset{G}}

\title{Hybrid Compositional Reasoning for Reactive Synthesis\\ from Finite-Horizon Specifications}
\author{Suguman Bansal$^\dagger$, Yong Li$^{\ddagger*}$, Lucas M. Tabajara$^\dagger$, Moshe Y. Vardi$^\dagger$\\
	$^\dagger$Department of Computer Science, Rice University\\
	$^\ddagger$State Key Laboratory of Computer Science, \\ Institute of Software, Chinese Academy of Sciences\\
	$^*$ University of Chinese Academy of Sciences}

\begin{document}

\maketitle

\begin{abstract}

$\ltlf$ synthesis is the automated construction of a reactive system from a high-level description, expressed in $\ltlf$, of its finite-horizon behavior. So far, the conversion of $\ltlf$ formulas to deterministic finite-state automata (DFAs) has been identified as the primary bottleneck to the scalabity of synthesis.
Recent investigations have also shown that the size of the DFA state space plays a critical role in synthesis as well.

Therefore, effective resolution of the bottleneck for synthesis requires the conversion to be time and memory performant, and prevent state-space explosion. Current conversion approaches, however, which are based either on explicit-state representation or symbolic-state representation, fail to address these necessities adequately at scale: Explicit-state approaches generate minimal DFA but are slow due to expensive DFA minimization. Symbolic-state representations can be succinct, but due to the lack of DFA minimization they  generate such large state spaces that even their symbolic representations cannot compensate for the blow-up. 

This work proposes a {\em hybrid} representation approach for the conversion. Our approach utilizes both  explicit and symbolic representations of the state-space, and effectively leverages their complementary strengths. In doing so, we offer an $\ltlf$ to DFA conversion technique that addresses all three necessities, hence resolving the bottleneck. A comprehensive empirical evaluation on conversion and synthesis benchmarks supports the merits of our hybrid approach.

\end{abstract}

\section{Introduction}
\label{Sec:Intro}

Reactive synthesis is the automated construction, from a high-level description of its desired behavior, of a reactive system that continuously interacts with an uncontrollable external environment~\cite{church1957applications}. This declarative paradigm holds the promise of simplifying the task of designing provably correct reactive systems.

This work looks into the development of reactive synthesis from specifications in {Linear Temporal Logic over finite traces} ($\ltlf$), or \emph{$\ltlf$ synthesis}, for short. $\ltlf$ is a specification language that expresses rich and complex temporal behaviors over a {\em finite} time horizon~\cite{de2013linear}. This formalism has found application in specifying task plans in robotics~\cite{he2017reactive,lahijanian2015time}, safety-critical objectives~\cite{zhu2017symbolic}, business processes~\cite{PesicBA10}, and the like. 

Seminal results have established that $\ltlf$ synthesis is 2$\mathsf{EXPTIME}$-complete~\cite{de2015synthesis}. Since then, several  undertakings have led to algorithmic solutions for  synthesis~\cite{de2015synthesis,camacho2018finite}. The current state-of-the-art reduces  synthesis to a reachability game played on a  deterministic finite-state automaton, or DFA~\cite{zhu2017ltlfsymbolic}. The DFA is obtained by converting the input $\ltlf$ specification into a DFA that recognizes the same language. This conversion has been identified as
a primary scalability bottleneck in synthesis \cite{zhu2017ltlfsymbolic}.
This is not surprising as the DFA is known to be double-exponential in the size of the specification in the worst case~\cite{kupferman1999model}. In order to be effective for synthesis the conversion must, in addition to being time and memory performant, also prevent state-space explosion, as recent investigations have discovered that the efficiency of solving the game on a DFA is strongly affected by the size of the state space~\cite{tabajara2019partition}. This work contributes towards the development of  $\ltlf$-to-DFA conversion techniques that are aimed at advancing the scalability of $\ltlf$ synthesis. 

Prior works on $\ltlf$-to-DFA conversion have led to two contrasting algorithmic approaches. In the first approach \cite{zhu2017ltlfsymbolic}, the state-space of the DFA is represented explicitly, the construction is syntax driven, and the DFA is aggressively minimized. This approach first converts $\ltlf$ to an equivalent first-order-logic formula and then constructs a DFA for this formula using the $\mona$ tool~\cite{henriksen1995mona}. The $\mona$ algorithm first produces the binary syntax tree of the specification, then traverses the tree bottom-up while constructing the {\em minimal} DFA at each node. Consequently, it constructs the final DFA at the root of the tree in its canonical minimal form. Aggressive minimization can often prevent state-space explosion, as for many specifications arising from real-life situations the minimal DFAs are rarely more than exponential in the size of the specification, as opposed to double exponential~\cite{tabakov2012optimized}.
Yet, an exponential DFA might still be too large if the set of states is represented explicitly, and the overhead caused by aggressive DFA minimization grows rapidly with specification size.

%SB
The second approach, inspired by~\cite{tabajara2019partition}, represents the DFA state space symbolically, uses a compositional construction, and avoids minimizing the DFAs.
In compositional constructions,  the specification is decomposed into multiple smaller sub-specifications for which explicit DFA conversion is tractable. These intermediate DFAs are then composed to get the final DFA.
%, replacing the older intermediate DFAs with a new intermediate one, until the final DFA is obtained. 
The symbolic representation  encodes the state space of a DFA in a logarithmic number of bits,
potentially achieving a polynomial representation even for an exponential-sized DFA, depending on the complexity of the DFA's structure.
The existing compositional approach takes advantage of this by representing the intermediate DFAs symbolically. In this case, the DFAs are composed by simply taking the symbolic product {\em without} performing minimization. 
The problem with this, however, is that each symbolic product results in a DFA with a larger state space than its minimal DFA, as no minimization is performed. When the number of symbolic products is large, the overhead in the size of the state space magnifies. 
Because of this, this approach ultimately produces a state space that is so enlarged that not even the succinct symbolic representation can compensate for the blow-up.

The key issue with both  approaches is that their critical operation is effective at small scale but becomes inhibitory at large scale. Explicit approaches aggressively perform minimization, which is efficient on small DFAs but expensive on larger ones. Meanwhile, symbolic approaches  perform symbolic products without minimization. While few symbolic products are manageable, too many products may lead to a large blow-up in the size of the state space.

This work proposes a novel compositional approach that is able to overcome the drawbacks of both existing approaches.
Our approach utilizes a {\em hybrid} state-space representation, i.e., at different times it uses both the explicit and symbolic state representations for the intermediate DFAs. The core idea is to use explicit-state representation for the intermediate DFAs as long as minimization is not prohibitively expensive, and to switch over to symbolic state representation as soon as that occurs. This way, our hybrid-representation approach applies explicit state representation to small DFAs, and also delays the point at which switch-over to symbolic representation occurs, thus ensuring that fewer symbolic products have to be performed to generate the final DFA.  Therefore, by finding a balance between the two representations, our hybrid appoach is able to extract their benefits and mitigate their weaknesses.

We have implemented our $\ltlf$-to-DFA conversion algorithm, and its extension to $\ltlf$ synthesis via reachability games, in tools called $\lisa$ and $\mathsf{LisaSynt}$, respectively. 
A comprehensive empirical analysis reveals the merits of the proposed hybrid compositional approach on both DFA conversion and $\ltlf$ synthesis, as each tool outperforms the current state-of-the-art in runtime and memory consumption. In addition, the DFAs generated from $\lisa$ have size comparable to the minimal DFA and significantly smaller than those obtained from pure symbolic-state methods.

\section{Preliminaries}
\label{sec:preliminaries}

%We begin by setting up necessary notations and definitions.

\subsection{Linear Temporal Logic over finite traces}

{Linear Temporal Logic over finite traces} ($\ltlf$)~\cite{baier2006planning,de2013linear} extends propositional logic with finite-horizon temporal operators. In effect, $\ltlf$  is a variant of $\ltl$~\cite{pnueli1977temporal} that is interpreted over a finite rather than infinite trace. The syntax of an $\ltlf$ formula over a finite set of propositions $\ap$ is identical to $\ltl$, and defined as $\varphi:= a \in \ap \mid \ltlfNeg \varphi \mid \varphi \land \varphi \mid \varphi \lor \varphi \mid \ltlfX \varphi \mid \varphi \ltlfU \varphi \mid \ltlfF \varphi \mid \ltlfG \varphi$. 
Here $\ltlfX$ (Next), $\ltlfU$ (Until), $\ltlfF$ (Eventually), $\ltlfG$ (Always) are temporal operators.  
The semantics of $\ltlf$ can be found in~\cite{de2013linear}. 
W.l.o.g., we assume that every $\ltlf$ formula $\varphi$ is written as a conjunction of $\ltlf$ subformulas i.e. $\varphi = \bigwedge_{i=1}^n \varphi_i$.
The language of an $\ltlf$ formula $\varphi$, denoted by $\L(\varphi)$, is the set of finite words over $2^{\ap}$  that satisfy $\varphi$.

$\ltlf$ synthesis is formally defined as follows:

\begin{definition}[$\ltlf$ synthesis]
\label{def:ltlf-synthesis}

Let $\varphi$ be an $\ltlf$ formula over $\ap = \apI \cup \apO$ where the set of input variables $\apI$ and output variables $\apO$ are two disjoint sets of propositions.
%$\apI$ is the set of input variables and $\apO$ is the set of output variables.
We say $\varphi$ is realizable if there exists a strategy $\gamma: (2^{\apI})^{+} \rightarrow 2^{\apO}$ such that for every infinite sequence $\lambda = I_{0}, I_{1}, \cdots \in (2^{\apI})^{\omega}$ of interpretations over $\apI$, there exists $m \geq 0$ such that $\rho = (I_{0} \cup \gamma(I_{0})), (I_{1} \cup \gamma(I_{0}, I_{1})), \cdots, (I_{m} \cup \gamma(I_{0}, \cdots, I_{m}))$ satisfies $\varphi$.
The problem of $\ltlf$ synthesis is to decide whether a given $\varphi$ is realizable and to construct such a strategy if so. 
\end{definition}

Intuitively, $\ltlf$ synthesis can be perceived as  a game between an external environment and the desired system that take turns to assign values to input and output propositions, respectively. The system responds to the environment inputs using the strategy $\gamma$. The game is won by the system if its strategy is able to guarantee that the resultant input-output sequence will satisfy formula $\varphi$ after a finite number of turns. In our formulation of $\ltlf$ synthesis, like in~\cite{tabajara2019partition}, the environment plays first. Alternatively, the system may play first~\cite{zhu2017ltlfsymbolic}. 
Solving the alternative formulation requires only slight changes to the algorithm presented in (\textsection~\ref{sec:synthesis}). We 
adhere to the formulation in Definition~\ref{def:ltlf-synthesis}  in this paper as our benchmarks assume that formulation and all tools being compared support it.

\subsection{DFA and its representations}
A {\em deterministic finite automaton} (DFA)~\cite{thomas2002automata} is a tuple $\dfa = ( \alphabet, \states , \init, \trans, \finals )$ where
$\alphabet$ is a finite set of symbols (called an alphabet),
$\states$ is a finite set of states,
$\init \in \states$ is the initial state,
$\finals \subseteq \states$ is the set of accepting states and
$\trans : \states \times \alphabet \rightarrow \states $ is the transition function. A finite word $w = w_0\dots w_n \in \Sigma^*$ has a {\em run} $\rho = s_0\dots s_{n+1} \in S^{+}$ in $\dfa$ if for all $i\in \{0,\dots n\}$ we have that $s_{i+1} = \trans(s_i, w_i)$ and $s_0 = \init$. A run $\rho = s_0\dots s_{n+1}$ is an {\em accepting run} in $\dfa$ if $s_{n+1} \in F$.
A word $w$ is in the language of  $\dfa$, $\L(\dfa)$, if $w$ has an accepting run in $\dfa$.
A DFA is said to be {\em minimal} if the language represented by that DFA cannot be represented by another DFA with fewer states.
%Every DFA has a canonical minimal DFA which accepts the same language but with the fewest number of states.

Every $\ltlf$ formula $\varphi$ over $\ap$ can be converted into a DFA $\dfa$ with alphabet $\alphabet = 2^\ap$~\cite{de2013linear} such that $\L(\dfa) = \L(\varphi)$.
%A DFA $\dfa$ can be compactly represented {\em semi-symbolically} by encoding the alphabet $\Sigma$ using propositions i.e. $\Sigma = 2^{\ap}$, where  $\ap$ is a finite set of propositions.
If this DFA is constructed in a form that explicitly enumerates all DFA states, we call it an \emph{explicit-state} representation. A DFA over the alphabet $\alphabet = 2^\ap$ can also be compactly represented \emph{symbolically}, by also encoding the state space using a logarithmic number of propositions. The \emph{symbolic-state} representation of a DFA $\dfa = (2^\ap, \states , \init, \trans, \finals )$ is a tuple $\sdfa = (\sinit(\ZVar), \strans(\ZVar, \ap, \ZVar'), \sfinals(\ZVar))$. In this representation, $\ZVar = \{z_1,\dots z_n\}$ are propositions encoding the state space $\states$, with $n = \lceil \log |\states| \rceil$, and their primed counterparts $\ZVar' =  \{z'_1,\dots z'_n\}$ encode the next state.
Each state $s \in S$ corresponds to an interpretation $Z \in 2^\ZVar$ over propositions $\ZVar$. When representing the next state of the transition function, the same encoding is used for an interpretation $Z'$ over $\ZVar'$.
Then, $\sinit$, $\strans$ and $\sfinals$ are Boolean formulas representing $\init$, $\trans$ and $\finals$, respectively. $\sinit(\ZVar)$ is satisfied only by the interpretation of the initial state $\init$ over $\ZVar$. $\strans(\ZVar, \ap, \ZVar')$ is satisfied by interpretations $Z \in 2^\ZVar$, $P \in 2^\ap$ and $Z' \in 2^{\ZVar'}$ iff $\trans(s, P) = s'$, where $s$ and $s'$ are the states corresponding to $Z$ and $Z'$.
Lastly, $\sfinals(\ZVar)$ is satisfied by the interpretation $Z$ over $\ZVar$ corresponding to state $s\in \states$ iff $s \in \finals$.
%For instance, assume that $\states = \{0, \ldots, n - 1\}$ for some integer $n$.Therefore a state $s \in \states$ can be represented by its binary encoding over the state variables $\ZVar$. Let $n = 4$ and $s = 2$. Then $\ZVar = \{z_1, z_2\}$, as every state can be encoded in two bits, and the binary encoding of $s$ is $(z_1 = 0, z_2 = 1)$ where $z_{2}$ is the most significant bit.
The intersection of two DFAs $\dfa_1=(\sinit_1, \strans_1, \sfinals_1)$ and $\dfa_2 = (\sinit_2, \strans_2, \sfinals_2)$, denoted $\dfa_1 \wedge \dfa_2$,   is given by $(\sinit_1\wedge \sinit_2, \strans_1 \wedge \strans_2, \sfinals_1 \wedge \sfinals_2)$. In this paper, all Boolean formulas, including $\sinit$, $\strans$ and $\sfinals$ of a symbolic DFA, will be encoded using {\em Reduced Ordered Binary Decision Diagrams} (BDDs)~\cite{bryant1986graph}.

\subsection{DFA game}
\label{subsec:dfa-game}

A \emph{DFA game} is a reachability game between two players, called the \emph{environment} and the \emph{system}, played over a DFA with alphabet $2^{\apI \cup \apO}$.
%That is, $\alphabet = 2^{\apI \cup \apO}$.
The environment player assigns values to the input variables $\apI$, while the system assigns values to the output variables $\apO$.
The DFA game starts at the initial state of the DFA.
At each round of the game, first the  environment chooses an assignment $I$ to the $\apI$ variables, and then the system will choose an assignment $O$ to the $\apO$ variables. % based on the current history of the inputs.
The combined assignment $I \cup O$ determines the unique state the game moves to according to the transition function of the DFA.
The system \emph{wins} the game if the game reaches an accepting state of the DFA.
{\em Solving a DFA game} corresponds to determining whether there exists a strategy for the system to always win the game.

DFA games are known to be solvable in polynomial time with respect to the number of states~\cite{mazala2002infinite}. The algorithm determines if the initial state is a {\em winning state}, i.e., a state that is either accepting or from which, for every assignment $I$ to the $\apI$ variables, the system can always choose an assignment $O$ to the $\apO$ variables that leads to a winning state. More details will be given in (\textsection~\ref{sec:synthesis}).
If the initial state is a winning state, then there exists a winning strategy that can be represented by a Mealy machine that
determines the output of the system given the current state and input. For more details, refer to~\cite{tabajara2019partition}.

\section{Related work}

\subsubsection{$\ltlf$ to DFA conversion}

There are two commonly used approaches for the conversion currently. 
In the current state-of-the art approach,
the $\ltlf$ formula is translated into  first-order logic over finite traces, and then converted into a DFA by $\mona$, a more general conversion tool from monadic second-order logic to DFA~\cite{henriksen1995mona}.
The first $\ltlf$ synthesis tool $\syft$ utilizes this method for DFA generation~\cite{zhu2017ltlfsymbolic}.

An alternative approach, used by the tool $\spot$~\cite{duret2016spot}, is to translate the $\ltlf$ formula into an $\ltl$ formula with equivalent semantics, convert this formula into a B\"{u}chi automaton~\cite{gerth1995simple}, and then transform this B\"{u}chi automaton into a DFA. 
Both approaches generate a DFA in explicit state-space representation.

\subsubsection{DFA vs. NFA}
NFAs are more general than DFAs. In fact, NFAs can be constructed from an $\ltlf$ formula in a single-exponential blow-up as opposed to the double-exponential blow-up incurred for DFA construction. 
Various approaches for $\ltlf$-to-NFA with single-exponential blow-up have been described such as 
\cite{baier2006planning,de2015synthesis}.
Yet, in practice, single exponential NFA conversion tools do not perform as well as DFA conversion tools.  
\cite{tabakov2012optimized} shows that minimal DFAs from $\ltlf$ formulas tend to be orders of magnitude smaller than their NFA counterparts constructed from implementations of the single-exponential algorithms.

\subsubsection{$\ltlf$ synthesis}
As aforementioned, current state-of-the-art tool $\syft$~\cite{zhu2017ltlfsymbolic} uses $\mona$ to construct an explicit-state DFA, then converts this DFA into a symbolic representation in order to solve the game using a symbolic fixed-point computation. The explicit-state DFA construction has been identified as the primary bottleneck to $\syft$ as the length of the formula increases. 
Therefore, recent attempts in synthesis have been made  to avoid the explicit DFA construction. We describe these attempts below.

A recent approach attempted to avoid the full construction by instead decomposing the specification into  conjuncts, then converting each conjunct to an individual DFA~\cite{tabajara2019partition}. Since these conjuncts are smaller formulas, their explicit-state DFAs can be constructed efficiently.
The smaller DFAs are then converted into a symbolic representation and the game is solved over this {\em decomposed symbolic representation}. 
While  
the construction was indeed more efficient in terms of time and memory, the resulting DFA had a much larger state space.  This severely decreased the performance of the game-solving algorithm, rendering a poorly scaling procedure for $\ltlf$ synthesis. 

In another attempt to avoid explicit DFA construction, \cite{camacho2018finite} first constructs an NFA from the formula and then reduces  synthesis to fully-observable nondeterministic (FOND) planning. The NFA is determinized  on-the-fly during the  planning phase.
Even here, the specification is decomposed into conjuncts, which are separately converted to NFAs and used to encode to FOND.
Despite the generalization to NFAs, in practice FOND-based methods rely on DFA conversion tools since they are more competitive than existing NFA construction tools that incur a single-exponential blow up.
Previous experiments suggest the FOND-based approach is complementary with the approach based on explicit DFA construction, each being able to solve instances that the other cannot.

\subsubsection{Compositional techniques in temporal  synthesis}

Both \cite{tabajara2019partition} and \cite{camacho2018finite} benefit from compositional techniques as they both  decompose the input formula into conjuncts before construction of the respective automata.  Application-specific decomposition has also been shown to lead to an orders-of-magnitude improvement in $\ltlf$ synthesis for robotics~\cite{he2019efficient}. 

A precedent for compositional techniques exists also in synthesis of $\ltl$ over infinite traces, including in several state-of-the-art tools such as $\strix$~\cite{meyer2018strix} and $\acaciaplus$~\cite{bohy2012acacia}. 
$\strix$ decomposes the formula semantically, i.e., it generates a subformula if it belongs to a restricted fragment of $\ltl$ such as safety $\ltl$ or co-safety $\ltl$. This way it benefits from constructing automaton using more efficient fragment-specific algorithms.
On the other hand, $\acaciaplus$ decomposes the formula into conjuncts, which are each solved as a  separate safety game. The final solution is obtained by composing solutions from the separate safety games.

\section{Hybrid compositional DFA generation}
\label{sec:hybrid}

This section describes the primary contribution of this work. We present a novel compositional approach for $\ltlf$-to-DFA conversion. 
Our approach is based on using a hybrid-state representation, i.e., at different times it uses both explicit and symbolic-state representations for intermediate DFAs, as opposed to prior works in which only one of the two state-representations is used~\cite{zhu2017ltlfsymbolic,camacho2018finite,tabajara2019partition}. 
By diligent application of both representations, our hybrid approach is able to leverage their complementary strengths and render an algorithm that is not only competitive time- and memory-wise, but also  generates DFAs with small number of states.

Our compositional approach is comprised of two phases, called the {\em decomposition phase} and the {\em composition phase}.
In the decomposition phase, the input formula is first {\em decomposed} into smaller subformulas which are then converted into their equivalent DFAs using standard algorithms.
In the composition phase, the intermediate DFAs are {\em composed} to produce the final DFA.
We describe each phase for our hybrid approach in detail below.
The formal description of our algorithm has been deferred to the Appendix.  
 
\subsection{Decomposition phase}

The decomposition phase is the first step in our algorithm. This phase receives the $\ltlf$ formula $\varphi$ as input. 
We make an  assumption that the formula is given as the conjunction of multiple small $\ltlf$ subformulas, i.e., $\varphi = \bigwedge_{i=1}^n \varphi_i$ where each $\varphi_i$ is an $\ltlf$ formula in itself.
This assumption has been adopted as a standard practice in  synthesis domains as large specifications arising from applications tend to exhibit this form~\cite{filiot2010compositional,filiot2011antichains}. 

We interpret formula $\varphi$ as an $n$-ary syntax tree as opposed to a binary-tree. Consequently, the input formula  $\varphi = \bigwedge_{i=1}^n \varphi_i$ is decomposed into  $n$-subformulas $\varphi_1, \dots ,\varphi_n$.
Then each of these subformulas $\varphi_i$ is converted into its minimal DFA $\dfa_i$ in explicit-state representation.
This can be performed by an existing  tool~\cite{de2013linear,duret2016spot,henriksen1995mona,kupferman1999model}.  More advanced decomposition schemes could be adopted from~\cite{camacho2018finite}.

The  rationale behind this step is that existing explicit-state  tools are efficient in generating minimal DFA for small formulas. Since the subformulas are typically small in length, we are able to benefit from existing literature in this step. 

\subsection{Composition phase}

The composition phase receives the minimal DFAs $\dfa_i$ for subformulas $\varphi_i$ in the previous phase, which are represented with explicit states. Our goal in this phase is to construct a DFA corresponding to $\varphi$. In theory, this can be obtained by simply taking the intersection of DFAs $\dfa_i$.
In practice, the intersection of $n$ DFAs may lead to state-space explosion since DFA intersection is done by performing their product construction.
Therefore, the main focus of the composition phase is about how to efficiently construct the intersection without incurring state explosion. 
We discuss the salient features of our algorithm before describing it in detail. 

Briefly speaking, we perform the composition of DFAs in iterations. 
In each iteration, two DFAs are selected based on a {\em dynamic smallest-first heuristic}, which will be described below, and removed from the set.
A new DFA is formed by the product of the two selected DFAs. The new DFA will be minimized based on a {\em selective DFA heuristic}, which is also described below. The new DFA is then inserted back into the set. The new set is the input to the next iteration.
This continues until only one DFA remains, which is presented as the final DFA. 
In the following, we denote by $S_j$ the set of DFAs at the $j$-th iteration. Then $S_1 = \{\dfa_1, \dots, \dfa_{n}\}$, and $S_{n} = \{\dfa\}$ where $\dfa$ is the final output DFA.

In contrast to prior works which either use explicit states or symbolic states, the central feature of our algorithm is that it uses hybrid representation for DFAs, i.e., in different iterations all DFAs in $S_j$ are either represented in explicit- or symbolic-state form. Initially, 
all DFAs in $S_1$ are in explicit-state form.
This continues while the DFAs in $S_j$ have a small number of states, since the product and minimization of DFAs are efficient for small DFAs with explicit-state representation.
But as some DFAs in $S_j$ grow in size they require more memory and longer time to perform minimization. 
So, as soon as some DFA in $S_j$ reaches a large number of states,  all DFAs in $S_j$ are converted into symbolic-state representation, in which the DFAs are represented more succinctly.
By this time, hopefully, we are left with few DFAs in the set $S_j$.
Here onwards, all DFAs are represented in symbolic form until the end of the algorithm. Therefore, fewer DFAs in $S_j$ implies fewer symbolic products need to be performed, and hence limits the blow-up in state-space of the final DFA. 
 This way, our algorithm balances the strengths of both approaches, mitigates their individual drawbacks, and efficiently generates a small DFA, if not the minimal.

We now describe the two heuristics, namely 
{\em dynamic smallest-first composition of DFAs} and {\em selective DFA minimization} abbreviated to DSF and SDM, respectively. 

We first discuss DSF, which is used to decide which two DFAs should be composed in each iteration.
We observe that the order in which intersection of DFAs is performed does not affect the correctness of the final DFA since both Boolean conjunction and DFA intersection are associative and commutative operations.
In theory, we can design any criteria to select two DFAs to be composed at each iteration.
In practice, a careless choice of the two DFAs may produce an unnecessarily large intermediate DFA that causes the algorithm to fail at the composition phase due to the large memory footprint.  
Therefore, we aim to find an order that can  optimize time and space in the composition phase.
To help with that we use DSF, which as the name suggests chooses the smallest two DFAs in each iteration.
The DFAs with explicit states are chosen based on the number of states, while the DFAs with symbolic-state representation are chosen based on the number of nodes in the BDD representation of the transition function.
The intuition behind this heuristic is that if the algorithm would fail on the composition of the smallest two DFAs in that iteration, then it would probably fail on the composition of all other pairs of DFAs as well.

Next we discuss SDM, which decides when it is beneficial to perform DFA minimization after the intersection of DFAs in each iteration.
%In short, our heuristic SDM chooses to minimize DFAs with explicit-state representation only due to following observation.
DFA minimization has been proved to be critical to the performance of DFA generation in~\cite{henriksen1995mona} as it helps in maintaining a smaller number of states, which is also one of our critical parameters.
However, it is also an expensive operation.
Currently, the best known complexity for  minimization are $\mathcal{O}(n\log n)$ and  $\mathcal{O}(n^2)$ for explicit- and symbolic-state representations, respectively~\cite{hopcroft1971n,wimmer2006sigref}. Therefore, there is a tension between reducing the number of states and achieving efficiency.
To resolve this, we conducted an empirical study to evaluate the effect of  minimization. We observed that in most cases, minimization reduces the number of states by 2-3 times. While this is significant when the states are represented explicitly, in symbolic-state representation this leads to a reduction in 1-2 state variables only.
Therefore, we adhere to the SDM heuristic in which we minimize intermediate DFAs in explicit-state representation only.
There are two advantages to this. 
First, since minimization is performed on explicit-state representation only, by virtue of our algorithm design this occurs only when the DFAs are small. For these, the time spent in minimization is so low that it is worth maintaining minimal DFAs.
Second, by maintaining minimal DFAs in the explicit-form, the algorithm delays the switch over to symbolic form as the DFA sizes take longer to reach the thresholds. This leads to fewer symbolic products, which results in curbing the amount of blow-up in state-space. 

A semi-formal description of the steps of the algorithm are given below. The complete formal description has been deferred to the Appendix.   

\vspace{-0.1cm}
\subsubsection{Step 0. (Initial)}

We are given input formula $\varphi = \bigwedge_{i=1}^n \varphi_i$, and {\em switch-over threshold values} $\threshDFA, \threshProd > 0$.
The parameters $t_1$ and $t_2$ correspond
to the thresholds for the numbers of states in an individual DFA and in the product of two DFAs, respectively, to trigger the symbolic representation. 

\vspace{-0.1cm}
\subsubsection{Step 1. (Decomposition)}
Construct the minimal DFA $\dfa_i$ in explicit-state representation for all $i \in \{1,\dots, n\}$. 
Create the set $S_1 = \{\dfa_1,\dots, \dfa_n\}$.

\vspace{-0.1cm}
\subsubsection{Step 2. (Explicit-state composition)}

For $j \in \{1,\dots, n-1\}$, let $S_j = \{M_1,\dots, M_{n-j+1}\}$ be the set of DFAs in the $j$-th iteration. 

If $S_j$ has only one DFA, return that as the solution. 

Otherwise, if the DFAs in $S_j$  become too large, proceed to Step 3. Assume w.l.o.g. that $M_1$ and $M_2$ are the two DFAs chosen by the DSF heuristic. Let $|A|$ denote the number of states in a DFA $A$ represented in explicit-state form. 
If $\min(|M_1|, |M_2|) > \threshDFA$ or $(|M_1|\cdot |M_2|) > \threshProd$, move to Step 3.  Let $k$ be the iteration in which this occurs, i.e. when $j = k$.

Otherwise, as per SDM, construct DFA $P$ by minimization of $M_1 \cap M_2$. Then, create $S_{j+1} = \{P, M_3,\dots, M_n\}$ for the next iteration, and repeat Step 2.

\vspace{-0.1cm}
\subsubsection{Step 3. (Change state representation)}

Convert all DFAs in $S_k = \{M_1,\dots, M_{n-k+1}\}$ from explicit-state  to symbolic-state representation, and proceed to Step 4. Note that the state space of each DFA $M_i$ is encoded symbolically using a different set of state variables $\ZVar_i$, where all $\ZVar_i$ are disjoint. Since no more minimization occurs after this point, the total set of state variables $\ZVar = \ZVar_1 \cup \ldots \cup \ZVar_{n-k+1}$ defines the state space of the final DFA.

\vspace{-0.1cm}
\subsubsection{Step 4. (Symbolic-state composition)}
For $j \in \{k,\dots, n\}$, let $S_j = \{M_1,\dots, M_{n-i+1}\}$ be the set of DFAs in the $j$-th iteration. 

If $S_j$ has only one DFA, return that DFA as the solution. 

Otherwise, assume w.l.o.g. that $M_1$ and $M_2$ are the two DFAs chosen by the DSF  heuristic. Construct $P = M_1 \wedge M_2$. Recall that, since $M_1$ and $M_2$ are in symbolic form, we do not perform DFA minimization of $P$. Create $S_{i+1} = \{P, M_3,\dots, M_n\}$ for the next iteration, and repeat Step 4.

\section{$\ltlf$ synthesis}
\label{sec:synthesis}

$\ltlf$ synthesis can be reduced to solving a DFA game played on the DFA corresponding to the formula $\varphi$~\cite{de2015synthesis}. As explained in (\textsection~\ref{subsec:dfa-game}), this amounts to computing the set of winning states. If the initial state of the DFA is in this set, then the formula is realizable and a winning strategy can be constructed, otherwise not. 

In this section, we describe the winning set computation algorithm on a DFA game when its states are represented symbolically. This is a standard least-fixed point algorithm for reachability games with symbolic state space, and is similar to~\cite{zhu2017ltlfsymbolic,tabajara2019partition}. For sake of completion, we summarize the algorithm here. 

Let $\varphi$ be an $\ltlf$ formula over disjoint input and output propositions $\apI$ and $\apO$, respectively, and $\G = (\sinit(\ZVar), \strans(\ZVar, \ap, \ZVar'), \sfinals(\ZVar))$ be a symbolic DFA for $\varphi$. The DFA game is played on $\G$. In our case, this DFA is obtained from our hybrid compositional approach (\textsection~\ref{sec:hybrid}), which we assume is in symbolic form, since explicit-state outputs can easily be converted to symbolic form.

To compute the winning set of $\G$, we compute the least-fixed point of a Boolean formula $W_{i}(\ZVar)$ that denotes the set of states from which the system can win in at most $i$ steps of the DFA game. 
Initially, $W_{0}(\ZVar)$ is the set  $\sfinals(\ZVar)$ of accepting states.
At each iteration, the algorithm constructs $W_{i+1}(\ZVar)$ from $W_{i}(\ZVar)$ by adding those states from which the system is guaranteed to reach $W_{i}(\ZVar)$ in one step. Formally, 
\[
W_{i+1}(\ZVar) = W_{i}(\ZVar) \lor (\forall \apI. \exists \apO, \ZVar'. \strans(\ZVar, \apI \cup \apO, \ZVar') \land W_{i}(\ZVar'))
\]
where $W_{i}(\ZVar')$ can be obtained from $W_{i}(\ZVar)$ by substituting variables $\ZVar$ with $\ZVar'$.
 This continues until no more states can be added to $W_{i+1}(\ZVar)$, i.e., until it encounters the first index $i$ such that  $W_{i+1}(\ZVar) \equiv W_{i}(\ZVar)$.
 Since the number of states in the DFA is finite, the algorithm is guaranteed to terminate.
The initial state is present in the winning set, say $W_\mathsf{FP}(\ZVar)$, if $\sinit(\ZVar) \implies W_\mathsf{FP}(\ZVar)$ holds. Details on winning-strategy construction has been deferred to~\cite{tabajara2019partition}.

In this work, all Boolean formulas for $\G$ and all  $W_{i+1}(\ZVar)$  are represented as BDDs. All boolean operations, quantification and variable substitution are available in standard BDD libraries. Finally, $\equiv$ is a constant time operation in BDDs. 

The complexity of solving a DFA game is polynomial in the size of the state space.
Therefore, the efficiency of $\ltlf$ synthesis is heavily affected by the size of the constructed DFA. 
Therefore, as our hybrid compositional approach generates small (if not minimal) DFAs, these are suitable for synthesis, as witnessed also by our experimental evaluation.

\section{Experimental evaluation}
\label{sec:exp-eval}

The goal of the empirical analysis is to examine  the performance of our hybrid approach in $\ltlf$-to-DFA generation and  $\ltl$ synthesis against existing tools and approaches.% over a benchmark suite curated from several prior works. 

\subsection{Implementation details}
Our hybrid compositional $\ltlf$-to-DFA conversion procedure (\textsection~\ref{sec:hybrid}) has been implemented in a tool called $\lisa$.
$\lisa$ has been extended to $\lisasynt$  to perform $\ltlf$ synthesis using the winning strategy computation described in (\textsection~\ref{sec:synthesis}).

$\lisa$ takes an $\ltlf$ formula and switch-over thresholds  $\threshDFA$, $\threshProd$ as inputs,  and outputs a corresponding DFA with symbolic states. The output may not be minimal. 
For the same inputs, $\lisasynt$ internally invokes $\lisa$, solves the DFA game given by $\lisa$'s output, and returns whether the formula is realizable. If so, it can also return a winning strategy.

$\lisa$ and $\lisasynt$ have been written in C++. 
They employ $\mathsf{BuDDy}$~\cite{BuDDypakage} as their BDD library for the symbolic representations and operations on DFAs, and  take advantage of dynamic variable ordering for the BDDs.

To generate explicit-state minimal DFAs in the decomposition phase,
$\lisa$ uses $\spot$~\cite{duret2016spot} and the $\mona$-based method~\cite{henriksen1995mona}. 
It borrows the rich APIs from $\spot$ to conduct DFA intersection and minimization in the explicit-state composition phase.
%Per se, $\spot$  manipulates  automata over {\em infinite} words, so we write a wrapper that interprets DFAs as {\em deterministic co-safety automata} - a class of automata over infinte words for which  intersection and minimization closely correspond to the same on DFAs -  to interface  with $\spot$.
Per se, $\spot$ APIs are available for $\omega$-automata (automata over infinite words). In order to use the $\spot$ API for operations over DFAs, $\lisa$
stores intermediate explicit DFAs as a {\em weak deterministic B\"uchi automata }(wDBA)~\cite{dax2007mechanizing}.  
Intuitively, if the DFA accepts the language $\L$, then its wDBA accepts the language $\L\cdot(\{\ltlloop\})^\omega$, where $\ltlloop$ is a fresh variable not present in $\ap$. 
The wDBA can be constructed from the DFA for $\L$ by making the following changes (a) add a new state $\mathsf{sink}$, (b) for each accepting state in the DFA, add a transition from that state to $\mathsf{sink}$ on $\ltlloop$, (c) add a transition from  $\mathsf{sink}$ to itself on $\ltlloop$ , (d) make $\mathsf{sink}$ the only accepting state in the wDBA. This automaton accepts a word iff its run visits $\mathsf{sink}$ infinitely often. 
Since wDBA is an $\omega$-automaton, we use $\spot$ APIs for wDBAs to conduct intersection and minimization, both of which return a wDBA as output, in the similar complexity for those operations in a DFA~\cite{dax2007mechanizing,Kupferman18}.
Lastly, a wDBA for language $\L\cdot(\{\ltlloop\})^{\omega}$ can be easily converted back to a DFA for language $\L$.

\subsection{Design and setup for empirical evaluation \footnote{The source code of our tool is publicly available at \url{https://github.com/vardigroup/lisa}}\footnote{Figures are best viewed online in color.}}

The evaluation has been designed to compare
the performance of $\lisa$ and $\lisasynt$ to their respective existing tools and approaches. 
$\ltlf$-to-DFA conversion tools are compared on runtime, number of benchmarks solved, hardness of benchmarks solved (size of minimal DFA) and the number of state variables in the output DFA. $\ltlf$ synthesis tools are compared on runtime and the number of benchmarks solved.  
We conduct our experiments on a benchmark suite curated from prior works, spanning classes of realistic and synthetic benchmarks. In total, we have 454 benchmarks split into four classes: random conjunctions (400 cases)~\cite{zhu2017ltlfsymbolic}, single counters (20 cases), double counters (10 cases) and Nim games (24 cases)~\cite{tabajara2019partition}. More details on each class can be found in the Appendix.

\begin{table}[t]
\centering

\begin{tabular}{c|c|c|c}
\hline
\multicolumn{1}{|c|}{
\multirow{ 2}{*}{
\textbf{\begin{tabular}[c]{@{}c@{}}\# States in \\ the minimal \\ DFA\end{tabular}}
}
}
& \multicolumn{3}{l|}{\textbf{Number of benchmarks solved}}                                                                                                               \\ \cline{2-4} 
\multicolumn{1}{|c|}{}                                                                                                     & \textbf{\begin{tabular}[c]{@{}c@{}}Mona-\\ based\end{tabular}} & \textbf{\begin{tabular}[c]{@{}c@{}}Lisa-\\Explicit\end{tabular}} & \multicolumn{1}{c|}{\textbf{Lisa}} \\ \hline

\multicolumn{1}{c|}{$\geq$ 1K}   & 111 & 123 & 137 \\ \hline
\multicolumn{1}{c|}{$\geq$ 5K}   & 70 & 82 & 96 \\ \hline
\multicolumn{1}{c|}{$\geq$ 10K}   & 48 & 60 & 74 \\ \hline
\multicolumn{1}{c|}{$\geq$ 50K}   & 13 & 23 & 35 \\ \hline
\multicolumn{1}{c|}{$\geq$ 100K}   & 8 & 16 & 26 \\ \hline
\multicolumn{1}{c|}{$\geq$ 250K}   & 1 & 5 & 12 \\ \hline
\multicolumn{1}{c|}{$\geq$ 500K}   & 0 & 2 & 4 \\ \hline
\multicolumn{1}{c|}{$\geq$ 750K}   & 0 & 2 & 2 \\ \hline
\multicolumn{1}{c|}{Size unknown} & -- & -- & 21** \\ \hline
\multicolumn{1}{c|}{\textbf{Total solved}}   & \textbf{307} & \textbf{338} & \textbf{372} \\ \hline
\end{tabular}
\caption{DFA construction. Hardness of benchmarks is measured by the size of minimal DFA. **Note: There are 34 benchmarks that were solved only by $\lisa$. Of these, we were able to identify the size of the minimal DFA of 13 benchmarks using a symbolic DFA minimization algorithm~\cite{wimmer2006sigref}. The 21 cases with unknown size are those that could not be minimized even after 24hrs with 190GB.}
\label{table:size}
\vspace{-0.2cm}
\end{table}

\begin{figure}[t]
\centering
\includegraphics[width=0.42\textwidth, trim=1cm 0.8cm 1cm 1cm]{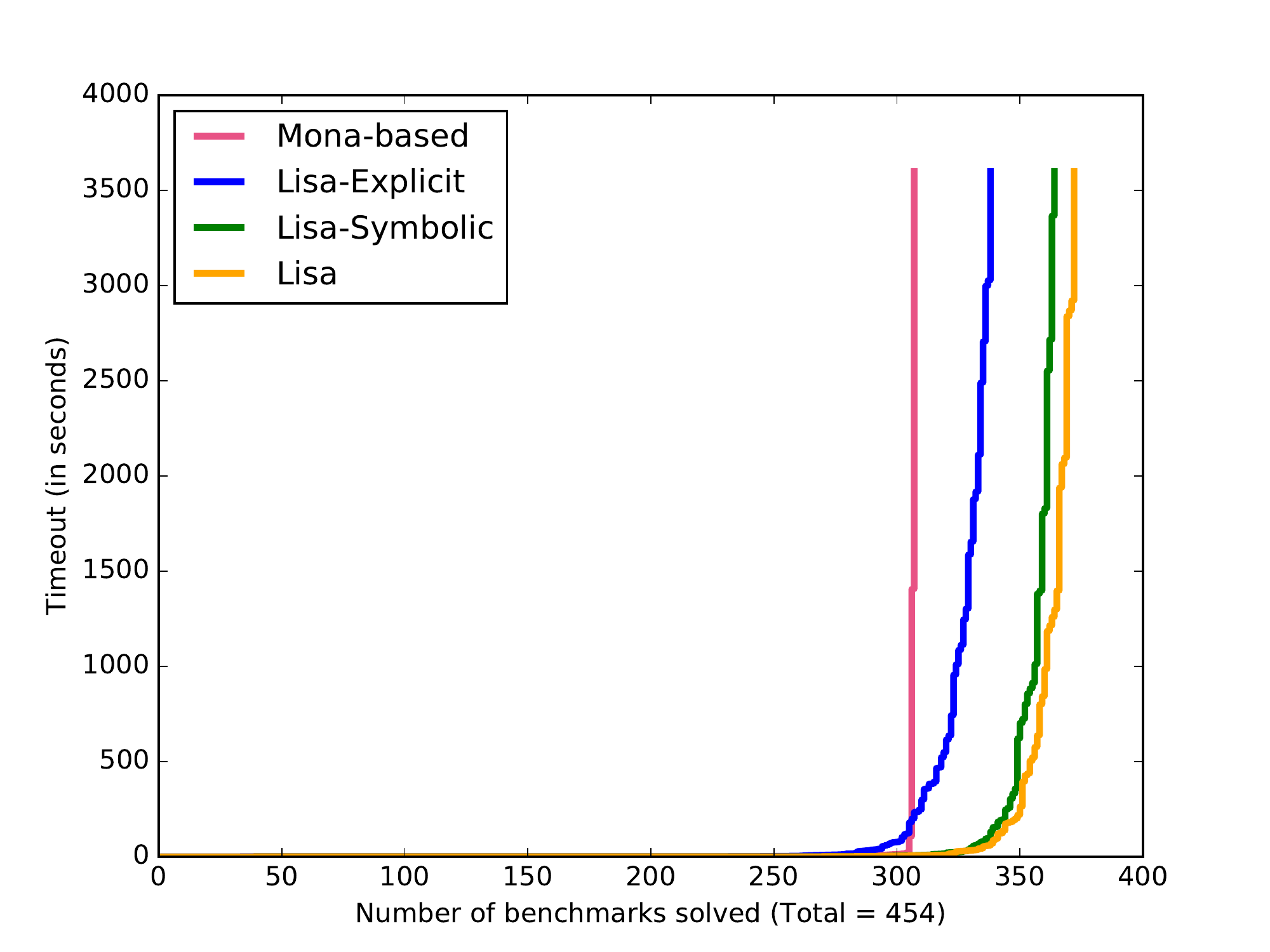}
%\vspace{0.5cm}
\caption{DFA construction. Cactus plot indicating number of benchmarks each tool can solve for a given timeout.}
\vspace{-0.5cm}
\label{Fig:cactusdfa}
\end{figure}

A good balance between explicit- and symbolic-representation of states is crucial to the performance of $\lisa$, i.e., it is crucial to carefully choose values of the switch-over thresholds $\threshDFA$ and $\threshProd$. Recall the switch is triggered if either the smallest minimal DFA has more than $\threshDFA$ states, or if the product of the number of states in the two smallest minimal DFAs is more than $\threshProd$. Intuitively, we want $\threshDFA$ to be large enough that the switch is not triggered too soon but small enough that conversion of all DFAs from explicit- to symbolic-state representation is not too expensive.  Threshold $\threshProd$ is closely related to how effective minimization is, and hence depends on the benchmark class. If the benchmark class is such that minimization reduces the DFA size by only 2-3 times, then we would set $\threshProd$ to be a low value. But if the class is such that minimization reduces DFA size by orders of magnitude, as it does for the Nim game class, we set $\threshProd$ to a higher value to take advantage of minimization. Currently, these are determined empirically. We set $\threshDFA$ and $\threshProd$ to 800 and 300000, respectively, for the Nim-game class and to 800 and 2500, respectively, for all other classes.

For experiments on $\ltlf$-to-DFA conversion, we compare $\lisa$ to 
the current-state-of-the-art  $\mona$-based method ~\cite{zhu2017ltlfsymbolic,camacho2018finite} and two other derivations of $\lisa$. Recall the $\mona$-based method is a syntax-driven, explicit-state based approach that returns minimal DFAs. The first derivation is  $\lisa$-$\mathsf{Explicit}$ which is adapted from $\lisa$ by setting $\threshDFA=\threshProd=\infty$. 
Therefore, it is a purely explicit-state compositional approach. Like the $\mona$-based method, it also generates the minimal DFA, but unlike the former it uses the smallest-first heuristic. 
The second derivation is 
 $\lisa$-$\mathsf{Symbolic}$, adapted from $\lisa$ by setting $\threshDFA=\threshProd=0$.
 This corresponds to the compositional, symbolic-state approach referred to in (\textsection~\ref{Sec:Intro}).

 For experiments on $\ltlf$ synthesis, we compared $\lisasynt$ to an enhanced version of $\syft$ (a tool that uses the $\mona$-based method for DFA conversion)~\cite{zhu2017ltlfsymbolic} that we call $\syft+$, and the partitioned approach from~\cite{tabajara2019partition}, referred to as $\partition$.
 $\syft+$ was created by enabling dynamic variable ordering in $\syft$. This was necessary for a fair comparison as $\syft$, unlike $\lisasynt$ and $\partition$, uses static variable ordering. We observed that $\syft+$ shows upto 75\% reduction in runtime compared to $\syft$. Note that $\partition$ uses the same symbolic-state approach as $\lisa$-$\mathsf{Symbolic}$ for constructing the DFAs, except that it skips the composition step, instead performing synthesis directly over the initial set of symbolic DFAs $S_1$.
 Ultimately, it still suffers from the state-space explosion, only in this case it happens during the
 winning-state computation.

All experiments were conducted on a single node of a high-performance cluster. Each node consists of four quad-core Intel-Xeon processor running at 2.6~GHz.
$\ltlf$-to-DFA conversion experiments were run for 1~hour with 8~GB each, $\ltlf$-synthesis experiments for 8~hours with 32~GB each.

\begin{figure}[t]
\centering
\includegraphics[width=0.42\textwidth, trim=1cm 0.8cm 1cm 1cm]{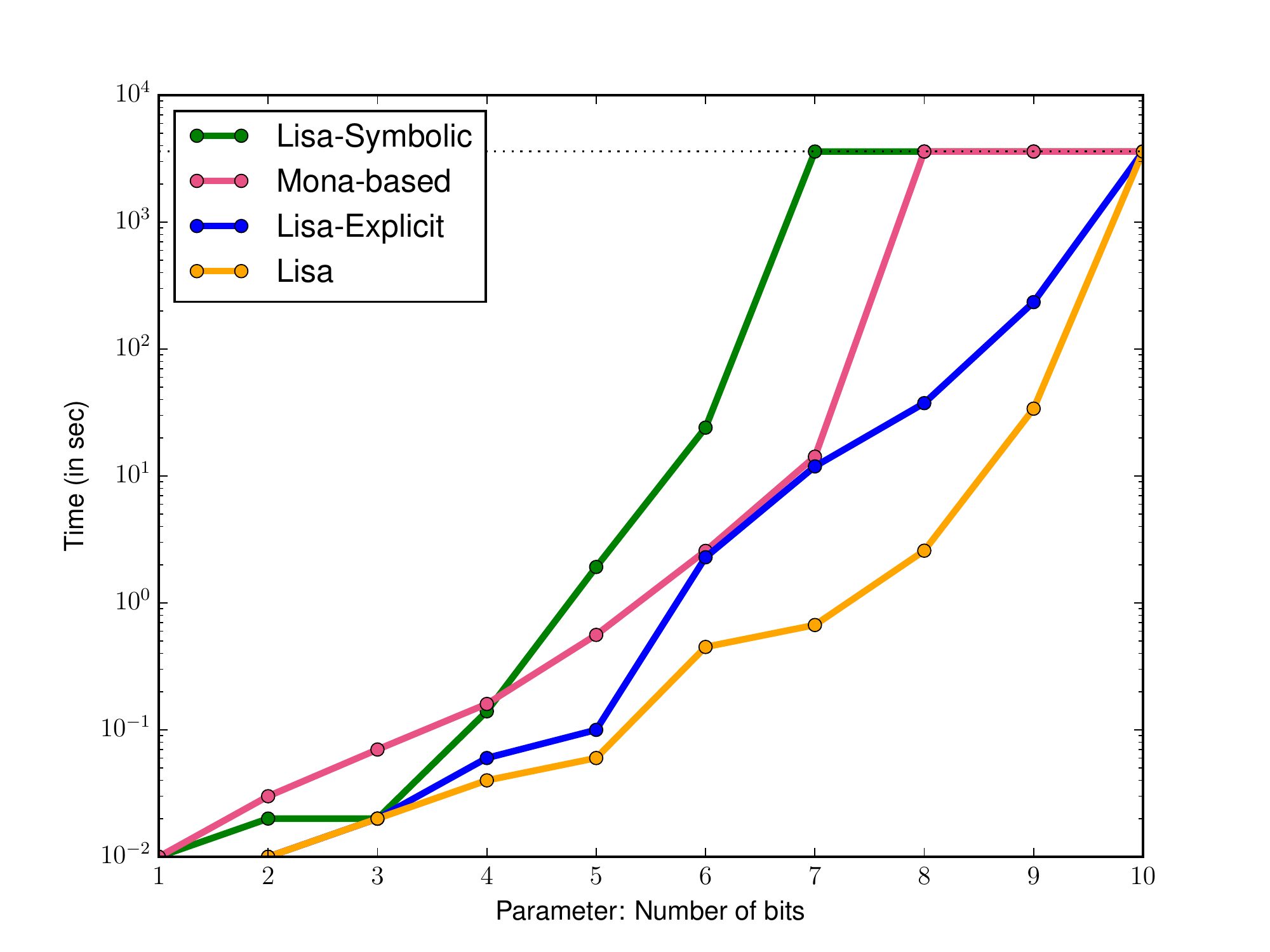}
%\vspace{0.5cm}
\caption{DFA construction. Runtime for double-counter benchmarks. Plots touching black line means time/memout.}
\vspace{-0.1cm}
\label{Fig:doubledfatime}
\end{figure}

\begin{figure}[!t]
\centering
\includegraphics[width=0.42\textwidth, trim=1cm 0.8cm 1cm 1cm]{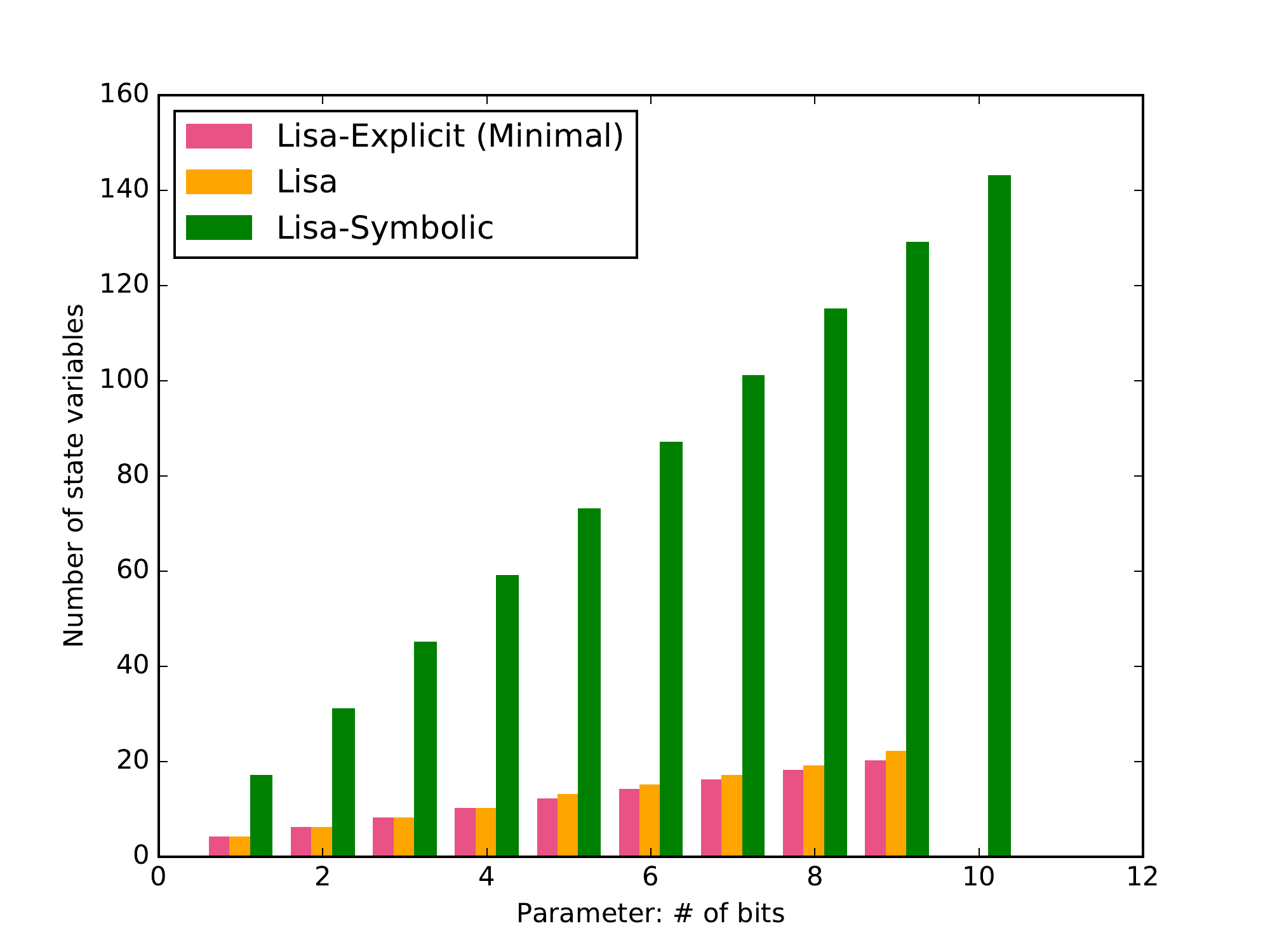}
%\vspace{0.5cm}
\caption{DFA construction. Number of variables needed to symbolically represent the  DFA's state-space for double-counter benchmarks. No bar indicates time/memout.}
\vspace{-0.5cm}
\label{Fig:doubledfavars}
\end{figure}

\subsection{Observations}

\subsubsection{$\lisa$ and $\lisa$-$\mathsf{explicit}$ scale better to larger benchmarks} than the $\mona$-based method, not just solving more total benchmarks but also being able to handle instances of larger scale (Table~\ref{table:size}).
Between $\lisa$-$\mathsf{Explicit}$ and $\mona$, the former is more consistent in solving benchmarks with large minimal DFAs due to the DSF heuristic that enables low memory consumption in intermediate stages.
Finally, $\lisa$ solves benchmarks with even larger minimal DFAsas it is designed to combine minimal DFAs of explicit state- and succinctness of symbolic-state representation to solve larger formulas.

\subsubsection{$\lisa$ is the most efficient tool among all four options.} This is clear from the cactus plot in Fig.~\ref{Fig:cactusdfa}. 
The plot may seem to indicate that $\mathsf{Lisa}$ only has a slight advantage over $\lisa$-$\mathsf{Symbolic}$. But, on closer inspection we observe that $\lisa$-$\mathsf{Symbolic}$ solves most random benchmarks but fares poorly on the realistic ones (see Fig~\ref{Fig:doubledfatime}). This is because they have more sub-specifications, resulting in a large number of symbolic products. The $\mona$-based method is still the fastest in generating small DFAs (fewer than 50K states) but memouts soon due to explicit-state representation of DFAs. Finally, $\lisa$-$\mathsf{Explicit}$ is a close second but does not scale as well as $\lisa$ due to  minimization on very large DFAs. $\lisa$  has been designed to overcome these deficiencies, and is supported by the current empirical evaluation as well.

\subsubsection{$\lisa$ mitigates state-space explosion.} Even though $\lisa$ may not generate the minimal DFAs, we observe that in most cases the state-space of the final DFA produced by $\lisa$ is one or two variables more than that of the minimal DFA. 
This is significantly lower than the number of state variables used by $\lisa$-$\mathsf{Symbolic}$ (Fig.~\ref{Fig:doubledfavars}). Note that $\lisa$-$\mathsf{Symbolic}$ fails to solve the double counter benchmarks for $i \geq 7$ (Fig~\ref{Fig:doubledfatime}). Yet we know the number of state variables immediately after Step 3 (\textsection~\ref{sec:hybrid}). 
Analyzing the benchmarks, we observed that they were split into 3-200 sub-formulas, yet only 1-3 symbolic products were conducted to construct the DFA.  
This demonstrates that our threshold-values are able to 
delay the switch-over to symbolic representations and reduce blow-up by the product.
This is why the DFAs generated by $\lisa$ have comparable sizes to the minimal DFAs.
An important future work, therefore, is to design mechanisms to determine the switch-over thresholds at runtime as opposed to relyng on user-expertise to  assign threshold values.

\subsubsection{$\lisa$'s small DFAs improve synthesis performance.}
\begin{figure}[t]

\centering
\includegraphics[width=0.42\textwidth, trim=1cm 0.8cm 1cm 1cm]{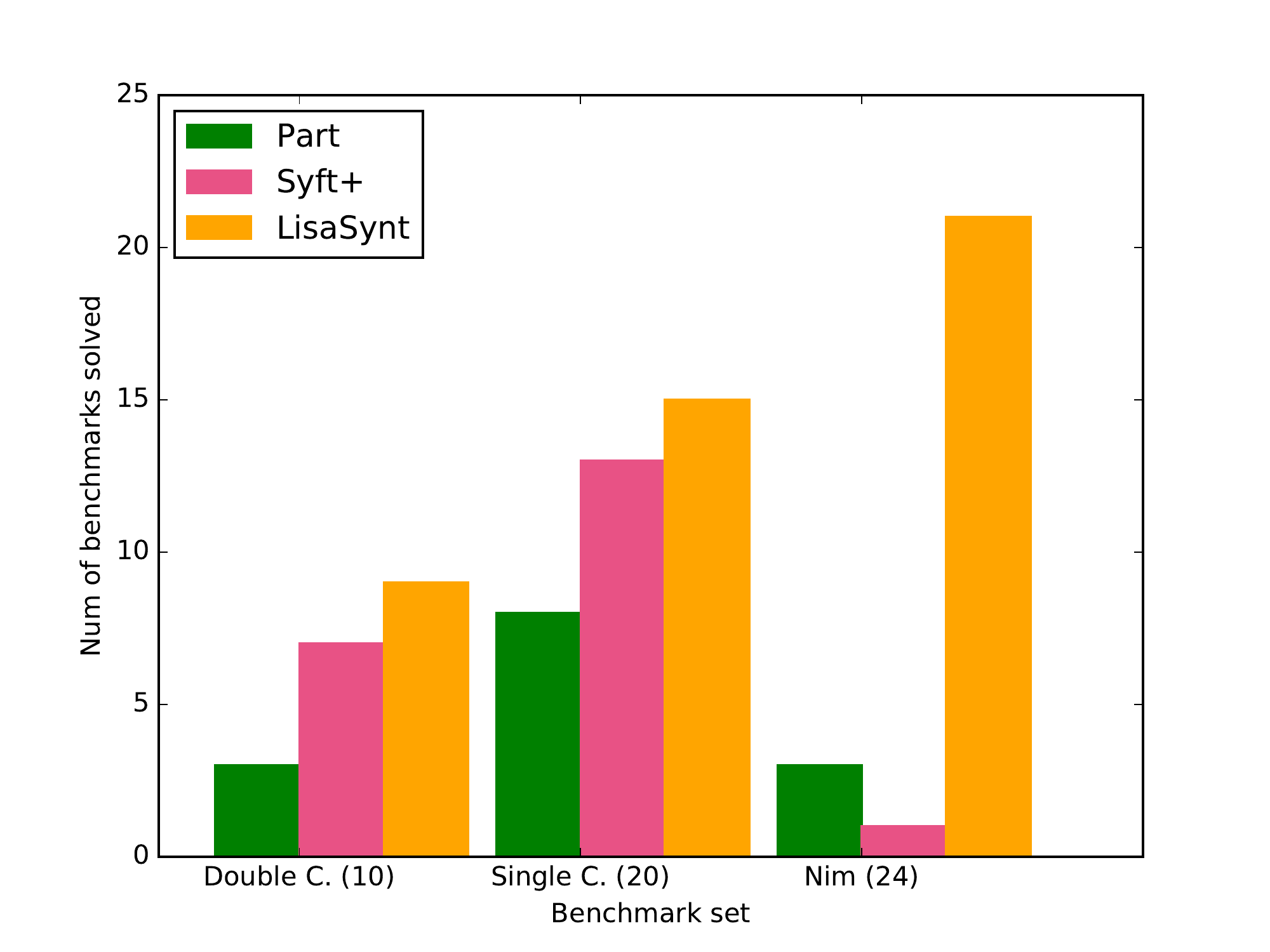}
\caption{Synthesis. Number of benchmarks synthesized from each non-random benchmark class.}
\label{Fig:doublesyntime}
\vspace{-0.0cm}
\end{figure}

\begin{figure}[t]

\centering
\includegraphics[width=0.42\textwidth, trim=1cm 0.8cm 1cm 1cm]{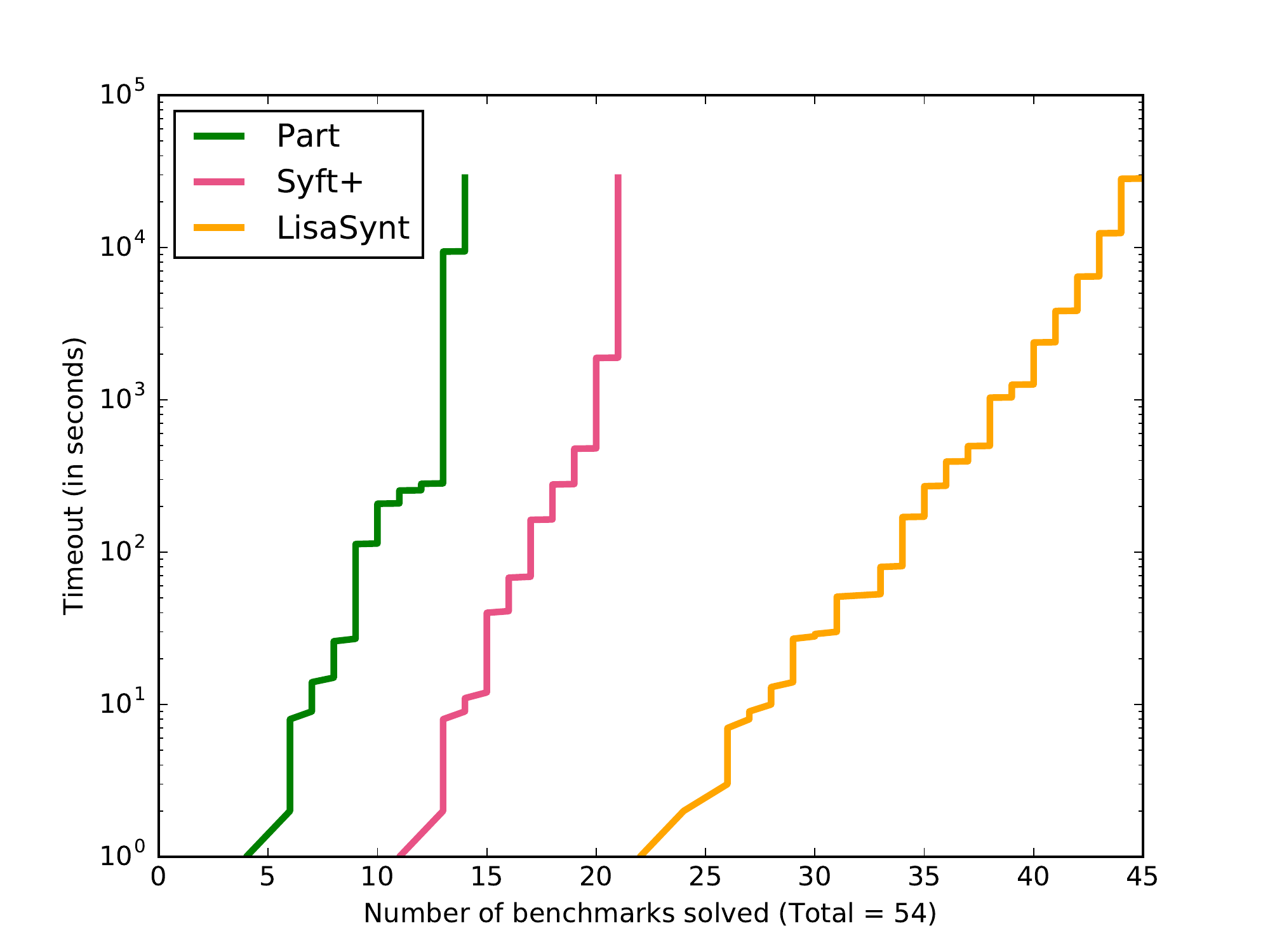}
\caption{Synthesis. Cactus plot  (non-random benchmarks).}
\label{Fig:cactussyn}
\vspace{-0.5cm}

\end{figure}

We evaluate for synthesis on non-random benchmarks only, i.e., sequential counters and nim games. We chose to disregard random benchmarks  as their winning set computation time is negligible, as in those benchmarks the fixed point is reached in 2-3 iteration irrespective of the DFA size. 
Figure~\ref{Fig:doublesyntime}-\ref{Fig:cactussyn} show that $\lisasynt$ solves most benchmarks and is the most efficient tool.
We observed that  $\syft$+ fails because $\mona$ memouts early, while $\partition$ suffers from state-space explosion. $\lisasynt$ is resilient to both as $\lisa$ consumes low memory by virtue of symbolic representation and small state space. 

The time consumed inside the winning set computation during synthesis depends on the number of iterations before the fixed-point is reached. Yet, so far not much focus has been given to optimizing this step as the DFAs generated so far have not been large enough for the number of iterations to become an issue.
With $\lisa$'s ability to construct large DFAs, we were able to observe that the single and double counter benchmarks can spend more than 90\% of the time  in the winning set computation, as the number of iterations is exponential in the number of bits (Appendix). 
This provides concrete evidence of the importance of investigating the development of faster algorithms for winning set computation to improve game-based synthesis.

\section{Concluding remarks}

This work tackles the primary bottleneck in $\ltlf$ synthesis- $\ltlf$ to DFA conversion.
The central problem addressed in this work is the efficient and scalable construction of DFAs with small state space from $\ltlf$ specifications, as a step to $\ltlf$ synthesis.
To the best of our knowledge, ours is the {\em first} {hybrid} approach for DFA construction. Our approach combines explicit- and symbolic-state representations in a manner that effectively leverages their strengths and alleviates their individual shortcomings. Our empirical evaluations on DFA conversion and $\ltlf$ synthesis on $\lisa$ and $\lisasynt$ outperform the current states of the art, and demonstrate the merit of our hybrid approach. This indicates promise to further develop and explore hybrid approaches for automaton generation for other specification languages as well, and encourages similar investigations into the other building blocks in synthesis algorithms. 

\section*{ Acknowledgments}
We thank A. Camacho, A. M. Wells and S. Zhu for their valuable inputs at different stages of the project. This work is partially supported by NSF grants IIS-1527668, CCF-1704883, IIS-1830549, the National Natural Science Foundation of China (Grant Nos. 61761136011, 61532019), the Guangdong Science and Technology Department (Grant No. 2018B010107004), and the Brazilian agency CNPq through the Ci\^{e}ncia Sem Fronteiras program.

\bibliographystyle{aaai}
\bibliography{refs}

\newpage

\section*{Appendix}

\section*{$\lisa$: Hybrid compositional DFA generation}

The DFA-construction algorithm used in $\lisa$  is described in (\textsection~\ref{sec:hybrid}). We give its 
psuedo-code here in Algorithm~\ref{Alg:lisa}.  

To recall, $\lisa$ is split into two phases. First is the decomposition phase, which splits the formula into smaller subformulas and converts each subformula into its minimal DFA in explicit-state representation (Line~\ref{algline:lisasym:decompose}-\ref{algline:lisasym:decomposeend}). 
Second is the composition phase (Line~\ref{algline:lisasym:decompose} onwards), which begins by performing explicit-state composition (Line~\ref{algline:lisasym:compose}-\ref{algline:lisasym:pop2}). When explicit-state composition becomes prohibitive (condition in Line~\ref{algline:lisasym:switchcondition}), all  DFAs are converted into symbolic-state representation (Line~\ref{algline:lissasym:heapbegin}-\ref{algline:lissasym:heapend}). Finally, after this, symbolic composition is conducted (Line~\ref{algline:lisasym:composebegin}-\ref{algline:lisasym:composeend}). 
Crucial heuristics adopted in the algorithm are dynamic smallest-first and selective DFA minimization. 

We implement the smallest-first heuristic using a priority queue, as using this data structure we can efficiently obtain the smallest elements in the collection. Priority queues $\dfasizeminheap$ and $\dfasymsizeminheap$ store DFAs in explicit and symbolic representations, respectively. The priority queues are implemented such that they give higher priority to ones with fewer number of states and fewer number of BDD nodes in the transition relation, respectively. 

 \begin{theorem}
\label{thrm:lisasymcorrectnes}
Let $\threshDFA, \threshProd > 0$ be  threshold values.  Given an $\ltlf$ formula $\varphi$, $\lisa(\varphi, \threshDFA, \threshProd)$ returns a DFA for $\varphi$ with symbolic state. 
\end{theorem}

%(\sinit_1, \strans_1, \sfinals_1)
\begin{algorithm}[t]
	\caption{ $\lisa(\varphi, \threshDFA, \threshProd)$\\
	\textsf{Input:} $\ltlf$ formula $\varphi =\bigwedge_{i=1}^n \varphi_i$, \\
	DFA size threshold values $\threshDFA, \threshProd>0$ \\
	\textsf{Output:} A DFA for $\varphi$ with symbolic states}
	\label{Alg:lisa}
	
	\begin{algorithmic}[1]
		
		 \STATE{\color{blue}// Decomposition phase}
		 
	    \STATE {$\splitformulalist \leftarrow \{\varphi_1,\dots,\varphi_n\}$
		\label{algline:lisasym:decompose}}
		
		\STATE{$\dfasizeminheap = \{\}$}
		\FOR{$\varphi_i\in \splitformulalist$}
		    \STATE{$D \leftarrow \mathsf{constructExplictDFA(\varphi_i)}$}
		    \STATE{$\dfasizeminheap.\push(D)$}\label{algline:lisasym:decomposeend}
		\ENDFOR
        
        \STATE{\color{blue}// Composition phase}
        
        \STATE{\color{blue}// Explicit-state composition}
        \label{algline:lisasym:compose}
        
        \STATE {$D_1 \leftarrow \dfasizeminheap.\pop()$}\label{algline:lisasym:pop1}
        \STATE {$D_2 \leftarrow \dfasizeminheap.\pop()$}

        \WHILE{$|D_1|< \threshDFA$ \text{ and } $|D_1|\cdot |D_2| < \threshProd$}\label{algline:lisasym:switchcondition}
        \STATE {$D_{temp} \leftarrow (D_1\cap D_2$)}
        \STATE {$D \leftarrow \explicitminimization(D_{temp})$}\label{lisa:algline:explicitminimization}
        \STATE {$\dfasizeminheap.\push(D)$}
        
        \IF{$\dfasizeminheap.\size()==1$}
            \STATE{$D \leftarrow \dfasizeminheap.\pop()$}
            \STATE{$(\sinit,\strans, \sfinals) \leftarrow \makesymbolic()$}
            \RETURN{ $(\sinit,\strans, \sfinals)$}
        \ENDIF
        
        \STATE {$D_1 \leftarrow \dfasizeminheap.\pop()$}
        \STATE {$D_2 \leftarrow \dfasizeminheap.\pop()$}\label{algline:lisasym:pop2}
        \ENDWHILE
        
        \STATE{\color{blue}// Change state representation}
        
        \STATE {$\dfasymsizeminheap = \{\}$ }\label{algline:lissasym:heapbegin}
        \WHILE{$\dfasizeminheap.\size() > 0$}
            \STATE{$D\leftarrow \dfasizeminheap.\pop()$ } 
            \STATE{$(\sinit,\strans, \sfinals) \leftarrow \makesymbolic(D)$}\label{algline:lisasym:makesym}
            \STATE{$\dfasymsizeminheap.\push((\sinit,\strans, \sfinals))$} \label{algline:lissasym:heapend}
        \ENDWHILE
        
        %\FOR{$D \in \dfalist$}
        %    \STATE {$\dfasymsizeminheap.\push(\makesymbolic(D))$}
        %\ENDFOR
        
        \STATE{\color{blue}// Begin symbolic composition}
        \WHILE{$\dfasymsizeminheap.\size()> 1$}
            \label{algline:lisasym:composebegin}
            \STATE {$(\sinit_1,\strans_1, \sfinals_1) \leftarrow \dfasymsizeminheap.\pop()$}
            \STATE {$(\sinit_2,\strans, \sfinals_2) \leftarrow \dfasymsizeminheap.\pop()$}
            \STATE {$(\sinit,\strans, \sfinals) \leftarrow (\sinit_1\wedge \sinit_2, \strans_1\wedge \strans_2, \sfinals_1\wedge \sfinals_2)$}
            \STATE {$\dfasymsizeminheap.\push((\sinit,\strans, \sfinals))$}
        \ENDWHILE
            \label{algline:lisasym:composeend}
        	
        \STATE{$(\sinit,\strans, \sfinals) \leftarrow \dfasymsizeminheap.\pop()$}
	    \RETURN {$(\sinit,\strans, \sfinals)$}
		
	\end{algorithmic}
\end{algorithm}

\section*{Benchmark descriptions}

We evaluate on a  set of 454 benchmarks split into four classes:

\subsubsection{Randomly generated.}

We adopt the random $\ltlf$ formula generation procedure from literature~\cite{zhu2017ltlfsymbolic,camacho2018finite}. For a length parameter $\ell$, it selects $\ell$ base cases from a pool of $\ltl$ benchmarks interpreted with $\ltlf$ semantics~\cite{jobstmann2006optimizations}, takes their conjunction and renames propositions so that they are shared across conjuncts.
 It must be noted that a large value of $\ell$ does not guarantee a larger minimal DFA. 
 
In our experiments, $\ell$ ranges from 3-10. For each $\ell$, we create 50 benchmarks, adding up to 400 random benchmarks. 

\subsubsection{Single counter/Double counters.}

These benchmarks represent games played over binary counters, and are parameterized by the number of bits $\ell$~\cite{tabajara2019partition}. These benchmarks can have either a single counter, which the system must increment when signaled by the environment, or two counters, one controlled by each player, where the goal of the system is to reach the value in the environment counter. In these cases, larger $\ell$ results in a larger minimal DFA. 

In our experiments $\ell$ ranges from 1-20 and 1-10 for single and double-counter benchmarks, respectively.

\subsubsection{Nim game.}

These benchmarks model a generalized version of the game of Nim~\cite{bouton1901nim} with $p$ heaps and $q$ tokens per heap, taken from~\cite{tabajara2019partition}. 

We create a total of 24 such benchmarks. 

\section*{Experimental evaluation}

\subsubsection{Optimizing winning-set computation is the next challenge in $\ltlf$ synthesis.}
\begin{figure}[t]

\centering
\includegraphics[width=0.42\textwidth, trim=1cm 0.8cm 1cm 1cm]{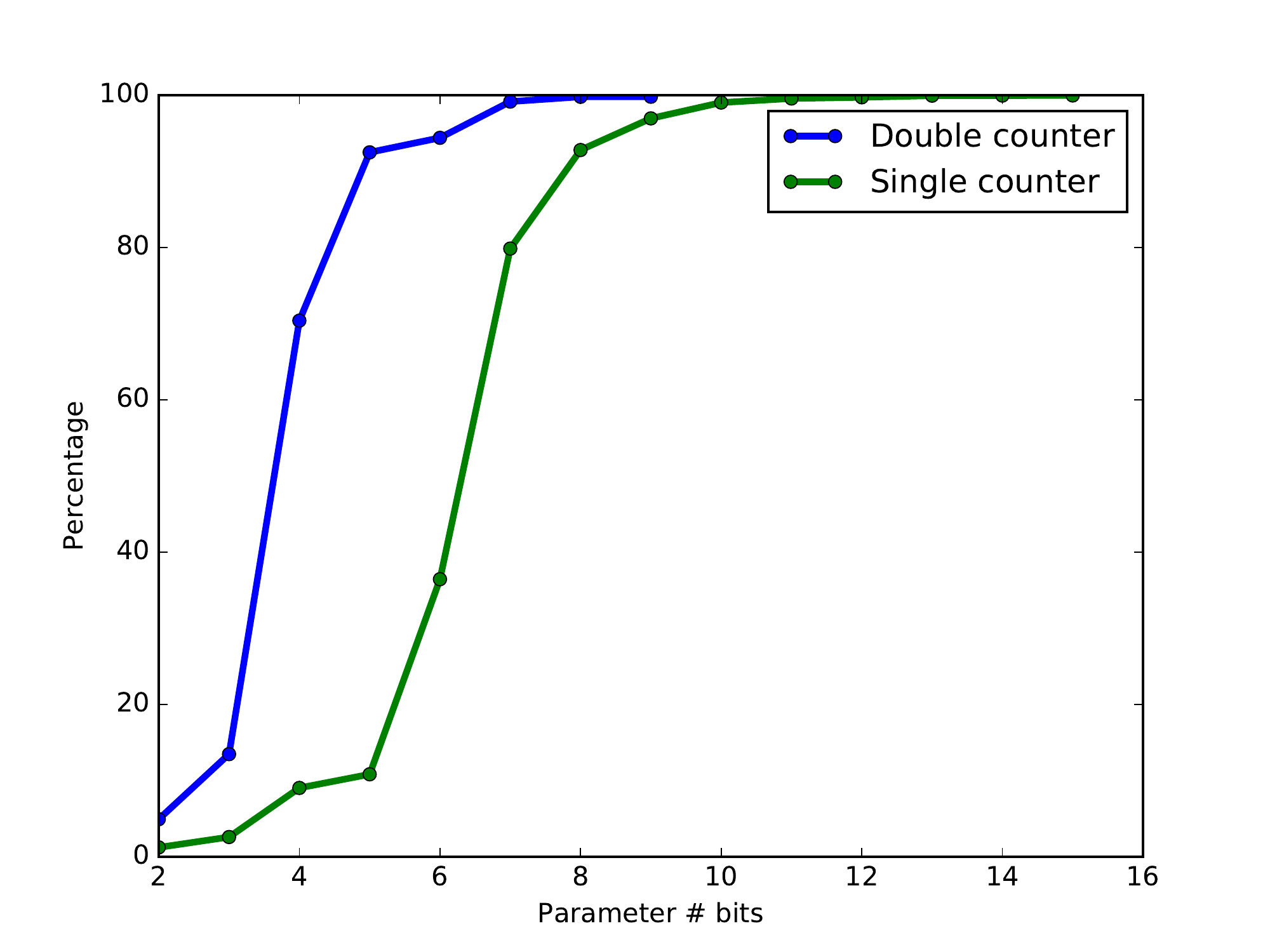}
\caption{Synthesis. Percentage of time spent in winning set computation for the sequential counters.}
\label{Fig:ratio}
\vspace{-0.5cm}

\end{figure}

Fig.~\ref{Fig:ratio} shows that most of the time spent in synthesis for the double-counter benchmarks was spent in the winning-set computation. In fact, for counters with $i$ bits, we observed and can show that the single- and double-counter benchmark will take $2^i$ and $2^{i+1}-2$ iterations to reach the fixed point in the winning-set computation. 

Winning-set computation takes almost 100\% of the time when $i=1$ since for those cases DFA construction requires less than one tenth of a millisecond to solve. As a consequence, winning-set computation takes very long in comparison. 

The plot in Fig.~\ref{Fig:ratio} is based on runtimes from $\lisasynt$. Although similar behavior was observed with $\syft$+, it solves fewer benchmarks, therefore we chose our tool to plot this graph. It must be noted that $\syft$ without dynamic variable ordering did not scale as far as $\syft+$.

% \usepackage{multirow}
\begin{table*}[]
\centering
\begin{tabular}{|c|c|c|c|c|c|c|}
\hline
\multirow{2}{*}{\textbf{Benchmark name}} & \multicolumn{3}{c|}{$\syft$+}                 & \multicolumn{3}{c|}{$\lisasynt$}             \\ \cline{2-7} 
                                         & DFA ($\mona$) & \textbf{WS} & \textbf{Total} & DFA ($\lisa$) & \textbf{WS} & \textbf{Total} \\ \hline
\multicolumn{7}{|c|}{\textbf{Single counters}}                                                                                         \\ \hline

counter-1 & 0.0845936 & 0.0063098 & 0.0909034 & 0.0 & 4.7624e-05 & 4.7624e-05    \\ \hline 
counter-2 & 0.0905504 & 0.0077225 & 0.0982729 & 0.01 & 0.000124862 & 0.010124862    \\ \hline 
counter-3 & 0.0831357 & 0.0196413 & 0.102777 & 0.01 & 0.000266569 & 0.010266569    \\ \hline 
counter-4 & 0.0818164 & 0.0125855 & 0.0944019 & 0.01 & 0.000994192 & 0.010994192    \\ \hline 
counter-5 & 0.113252 & 0.067858 & 0.18111 & 0.02 & 0.00242963 & 0.02242963    \\ \hline 
counter-6 & 0.104034 & 0.149737 & 0.253771 & 0.03 & 0.0172029 & 0.0472029    \\ \hline 
counter-7 & 0.163252 & 0.463409 & 0.626661 & 0.02 & 0.0793083 & 0.0993083    \\ \hline 
counter-8 & 0.120607 & 1.477873 & 1.59848 & 0.03 & 0.386659 & 0.416659    \\ \hline 
counter-9 & 0.260158 & 7.952932 & 8.21309 & 0.07 & 2.22855 & 2.29855    \\ \hline 
counter-10 & 0.383924 & 40.069576 & 40.4535 & 0.09 & 9.22808 & 9.31808    \\ \hline 
counter-11 & 0.853298 & 67.246602 & 68.0999 & 0.22 & 50.9045 & 51.1245    \\ \hline 
counter-12 & 1.83989 & 276.44511 & 278.285 & 0.75 & 270.575 & 271.325    \\ \hline 
counter-13 & 4.77399 & 474.14001 & 478.914 & 1.48 & 2378.66 & 2380.14    \\ \hline 
counter-14 & 11.19 & -- & -- & 3.25 & 6439.63 & 6442.88    \\ \hline 
counter-15 & -- & -- & -- & 9.93 & 28250.9 & 28260.83    \\ \hline 
counter-16 & -- & -- & -- & 43.01 & -- & --    \\ \hline 
counter-17 & -- & -- & -- & 174.04 & -- & --    \\ \hline 
counter-18 & -- & -- & -- & 191.31 & -- & --    \\ \hline 
counter-19 & -- & -- & -- & 2062.73 & -- & --    \\ \hline 
counter-20 & -- & -- & -- & 17499.8 & -- & --    \\ \hline

\multicolumn{7}{|c|}{\textbf{Double counters}}                                                                                         \\ \hline

counters-1 & 0.0372115 & 0.0088049 & 0.0460164 & 0.0 & 7.9756e-05 & 7.9756e-05   \\ \hline 
counters-2 & 0.0684606 & 0.0199104 & 0.088371 & 0.01 & 0.00051927 & 0.01051927   \\ \hline 
counters-3 & 0.140466 & 0.130129 & 0.270595 & 0.03 & 0.00467434 & 0.03467434   \\ \hline 
counters-4 & 0.206383 & 0.852017 & 1.0584 & 0.03 & 0.0713502 & 0.1013502   \\ \hline 
counters-5 & 0.488736 & 11.261164 & 11.7499 & 0.05 & 0.61636 & 0.66636   \\ \hline 
counters-6 & 2.71628 & 160.86572 & 163.582 & 0.44 & 7.42408 & 7.86408   \\ \hline 
counters-7 & 13.768 & 1871.902 & 1885.67 & 0.68 & 80.2925 & 80.9725   \\ \hline 
counters-8 & -- & -- & -- & 2.16 & 1032.24 & 1034.4   \\ \hline 
counters-9 & -- & -- & -- & 25.11 & 12372.5 & 12397.61   \\ \hline 
counters-10 & -- & -- & -- & 12885 & -- & --   \\ \hline

\multicolumn{7}{|c|}{\textbf{Nim benchmarks}}                                                                                          \\ \hline

nim-1-1 & 0.165076 & 0.010547 & 0.175623 & 0.08 & 4.133e-05 & 0.08004133    \\ \hline 
nim-1-2 & -- & -- & -- & 0.19 & 0.000140955 & 0.190140955    \\ \hline 
nim-1-3 & -- & -- & -- & 0.06 & 0.000197191 & 0.060197191    \\ \hline 
nim-1-4 & -- & -- & -- & 0.11 & 0.000382304 & 0.110382304    \\ \hline 
nim-1-5 & -- & -- & -- & 0.22 & 0.000372384 & 0.220372384    \\ \hline 
nim-1-6 & -- & -- & -- & 0.43 & 0.000530033 & 0.430530033    \\ \hline 
nim-1-7 & -- & -- & -- & 0.82 & 0.00065162 & 0.82065162    \\ \hline 
nim-1-8 & -- & -- & -- & 1.38 & 0.000906586 & 1.380906586    \\ \hline 
nim-2-1 & -- & -- & -- & 0.08 & 0.000329557 & 0.080329557    \\ \hline 
nim-2-2 & -- & -- & -- & 0.46 & 0.000856596 & 0.460856596    \\ \hline 
nim-2-3 & -- & -- & -- & 2.67 & 0.00185017 & 2.67185017    \\ \hline 
nim-2-4 & -- & -- & -- & 13.43 & 0.00650035 & 13.43650035    \\ \hline 
nim-2-5 & -- & -- & -- & 52.9 & 0.0162642 & 52.9162642    \\ \hline 
nim-2-6 & -- & -- & -- & 170.92 & 0.0330554 & 170.9530554    \\ \hline 
nim-2-7 & -- & -- & -- & 497.52 & 0.0535121 & 497.5735121    \\ \hline 
nim-2-8 & -- & -- & -- & 1257.86 & 0.106837 & 1257.966837    \\ \hline 
nim-3-1 & -- & -- & -- & 1.09 & 0.000209125 & 1.090209125    \\ \hline 
nim-3-2 & -- & -- & -- & 27.15 & 0.00700274 & 27.15700274    \\ \hline 
nim-3-3 & -- & -- & -- & 393.38 & 0.0396764 & 393.4196764    \\ \hline 
nim-3-4 & -- & -- & -- & 3820.42 & 10.8698 & 3831.2898    \\ \hline 
nim-4-1 & -- & -- & -- & 29.32 & 0.00661889 & 29.32661889    \\ \hline 
nim-4-2 & -- & -- & -- & -- & -- & --    \\ \hline 
nim-5-1 & -- & -- & -- & -- & -- & --    \\ \hline 
nim-5-2 & -- & -- & -- & -- & -- & --    \\ \hline

\end{tabular}
\caption{Runtime chart comparing runtimes of $\syft$+ and $\lisasynt$, and the time taken in DFA construction by $\mona$ and $\lisa$, respectively, inside the respective synthesis tool. Runtime are given in seconds. -- indicates time/memout. Timeout = $28800\sec$.}
\label{tab:runtime}
\end{table*}

\subsubsection{$\lisasynt$ outperforms the current state-of-the-art $\syft$+.}
This is clear from Table~\ref{tab:runtime}.
The main reason $\syft$+ fails to  solve a large number of benchmarks is that $\mona$ fails to generate the DFAs for larger inputs. For example, $\mona$ failed to construct the DFA for single counters and double counters for $i \geq 15$ and $i\geq 8$, where $i$ is the number of bits.
On the contrary, $\lisa$ is able to generate the DFAs for almost all benchmarks.

There are cases in Table~\ref{tab:runtime} where $\lisa$ has generated the DFA but $\lisasynt$ times-out. This is because the winning set computation did not terminate on those cases. Recall, the number of iterations for the counter benchmarks grows exponentially with the number of bits. These cases will be solved by $\lisasynt$ as long as enough time is given to conduct all iterations. 

Even for the benchmarks that are solved by both tools $\syft$+ and $\lisasynt$, $\lisasynt$ shows lower runtime. Note that for both tools the number of iterations taken to compute the winning set is the same. This may indicate that the time taken for winning set computation for each iteration in $\syft$+ takes longer that $\lisasynt$. However, currently we do not have concrete evidence to back-up this conjecture. We leave this to future work as it may also lead to a better understanding of how to improve the winning-set computation.

\end{document}